\documentclass[usenatbib]{mnras}

\usepackage[authoryear]{natbib}
\usepackage{fancyvrb}
\usepackage{amssymb}
\usepackage{amsmath}
\usepackage{cprotect}
\usepackage{setspace}
\usepackage{float}
\usepackage{graphicx}
\usepackage{color}
\usepackage{multirow,multicol}

\newcommand{\msol}{\mbox{M$_\odot \ $}}

\begin{document}

\VerbatimFootnotes

\title[Chemical evolution from RGB]{From K giants to G dwarfs: stellar lifetime effects on metallicity distributions derived from red giants}

\author[Manning \& Cole]{Ellen\ M. Manning,$^{1}$ Andrew\ A. Cole$^{1}$ \\
  \\
  $^1$ School of Physical Sciences, University of Tasmania, Private Bag 37, Hobart, Tasmania 7001, Australia}

 \maketitle

 \begin{abstract}

We examine the biases inherent to chemical abundance distributions when targets are selected from the red giant branch (RGB), using simulated giant branches created from isochrones. We find that even when stars are chosen from the entire colour range of RGB stars and over a broad range of magnitudes, the relative numbers of stars of different ages and metallicities, integrated over all stellar types, are not accurately represented in the giant branch sample. The result is that metallicity distribution functions derived from RGB star samples require a correction before they can be fit by chemical evolution models. We derive simple correction factors for over- and under-represented populations for the limiting cases of single-age populations with a broad range of metallicities and of continuous star formation at constant metallicity; an important general conclusion is that intermediate-age populations ($\approx$1--4~Gyr) are over-represented in RGB samples. We apply our models to the case of the Large Magellanic Cloud bar and show that the observed metallicity distribution underestimates the true number of metal-poor stars by more than 25\%; as a result, the inferred importance of gas flows in chemical evolution models could potentially be overestimated. The age- and metallicity-dependences of RGB lifetimes require careful modelling if they are not to lead to spurious conclusions about the chemical enrichment history of galaxies.

\end{abstract}

\begin{keywords}
stars: evolution --- stars: low-mass --- galaxies: evolution
\end{keywords}

\section{Introduction}
\label{sec:intro}

The buildup of chemical elements over time is a fundamental process whose details are a key part our understanding of the physics of the star-gas cycle in galaxies. To observationally measure the chemical evolution of the Milky Way (MW) in the solar neighbourhood, the classic approach has generally been to study the metallicity distribution function (MDF) of stars with Main Sequence lifetimes longer than the age of the Universe (M/M$_{\sun}$~$\lesssim$~0.8), so that the sample accurately represents the relative numbers of stars of all ages. An unbiased sample that can be directly compared to the predictions of chemical evolution models can thus be obtained \citep{vandenBergh62,Tinsley76}.  The problem with this is that the G and K dwarf stars which meet this criterion are very faint (M$_V$ $\gtrsim$ $+$5), and therefore hard to observe.  

The most easily observed abundances in external galaxies are those of the most luminous and massive stars and the gas phase, nebular abundances. Both of these provide a snapshot of the end result of billions of years of activity, introducing a high degree of model-dependence into conclusions about the the galactic chemical evolution. At older ages, red giants (spectral type K or early M) are the easiest type of star to observe for a very wide range of ages from the oldest stars in a galaxy to populations as young as $\approx$1~Gyr.  

In recent decades, and particularly since the advent of 8--10~m class telescopes, abundance measurements of hundreds to thousands of red giants in Milky Way satellites, isolated Local Group dwarfs, and in M31 and its satellites have been made \citep[e.g.][]{Suntzeff93,Tolstoy03,Tolstoy04,Cole05,Kirby11a,Kirby11b,Ho15,Ross15,Kirby17}.  A potential problem with this choice of target is that red giant branch (RGB) stars do not meet the immortality criterion that made dwarf stars the preferred targets for nearby stellar samples, potentially complicating the interpretation of the observed metallicity distribution functions.

It is well known that the position of an RGB star in the CMD depends strongly on the metallicity of the star \citep[e.g.][]{DaCosta90}; there is also a slight dependence on stellar age through the mass dependence of the radius of a star on the Hayashi track \citep{Hayashi62,Cole05}. The number of stars available to measure depends on the past star formation rate (SFR) of the galaxy and the initial mass function (IMF).  Where a metallicity distribution function of red giants is directly observed in a resolved stellar population, it is unlikely to represent the true proportion of heavy elements as a fraction of the total stellar mass of the galaxy, due to the finite lifetime of the red giants. Our purpose here is to investigate how these factors interact with the observational selection of a spectroscopic sample from a well-defined region of the CMD to create the observed MDF.  Other complications, such as the alteration of surface abundances due to heavy element diffusion or dredge-up, are beyond the scope of this paper.

In this paper, we use synthetic colour-magnitude diagrams (CMDs) to investigate the biases in observed MDFs at different ages and metallicities. Our simulations each approximate a simple stellar population with one metallicity and an age range over 1 Gyr. More complex systems can be built up by summing simple stellar populations of various age and metallicity, generating predictions for how many red giants are present as a function of magnitude, age, and metallicity in spectroscopic samples.  Section~\ref{sec:simulations} outlines our simulated CMD procedure.  Section~\ref{sec:results} details the results of our analysis.  In Section~\ref{sec:discussion} we discuss the implications of our results on current MDF models and in Section~\ref{sec:conclusions} we present our conclusions.

\section{Synthetic Red Giant Branches}
\label{sec:simulations}

\subsection{Red Giant Branch Lifetimes}
The lifetime of an RGB star is strongly dependent on its mass and metallicity, with low-metallicity stars evolving through the RGB phase more slowly than high-metallicity stars, and higher mass stars evolving more quickly than low-mass stars.  Because the mean mass of stars on the RGB decreases with time, this translates to an age-dependence.

In a galaxy with a very wide range of stellar ages, the RGB can include stars with a range of masses from $\sim$0.8~-~2\msol.  The potentially wide range of masses included in a sample must be treated with care to avoid biases caused by the initial mass function and evolutionary lifetime effects.  The IMF can be expressed as $\phi(m) \propto m^{-(1+x)}$ \citep[following the notation used by][]{Tinsley80}, and is the relative number or fraction of stars with initial masses between $(m, m+dm)$.  In real galaxies, the IMF is conflated with the SFR ($\psi(t)$), because the number of stars formed per unit time per unit mass interval is given by $\phi(m)\psi(t)$. 

For a given $\psi(t)$, the total mass of stars integrated along an isochrone, between the limits $(m, m+\Delta m)$ is

\begin{equation}
M_{\mathrm{tot}} = \int^{m+\Delta m}_m m\phi(m)dm.
\end{equation}

\begin{figure}
\includegraphics[width=\linewidth]{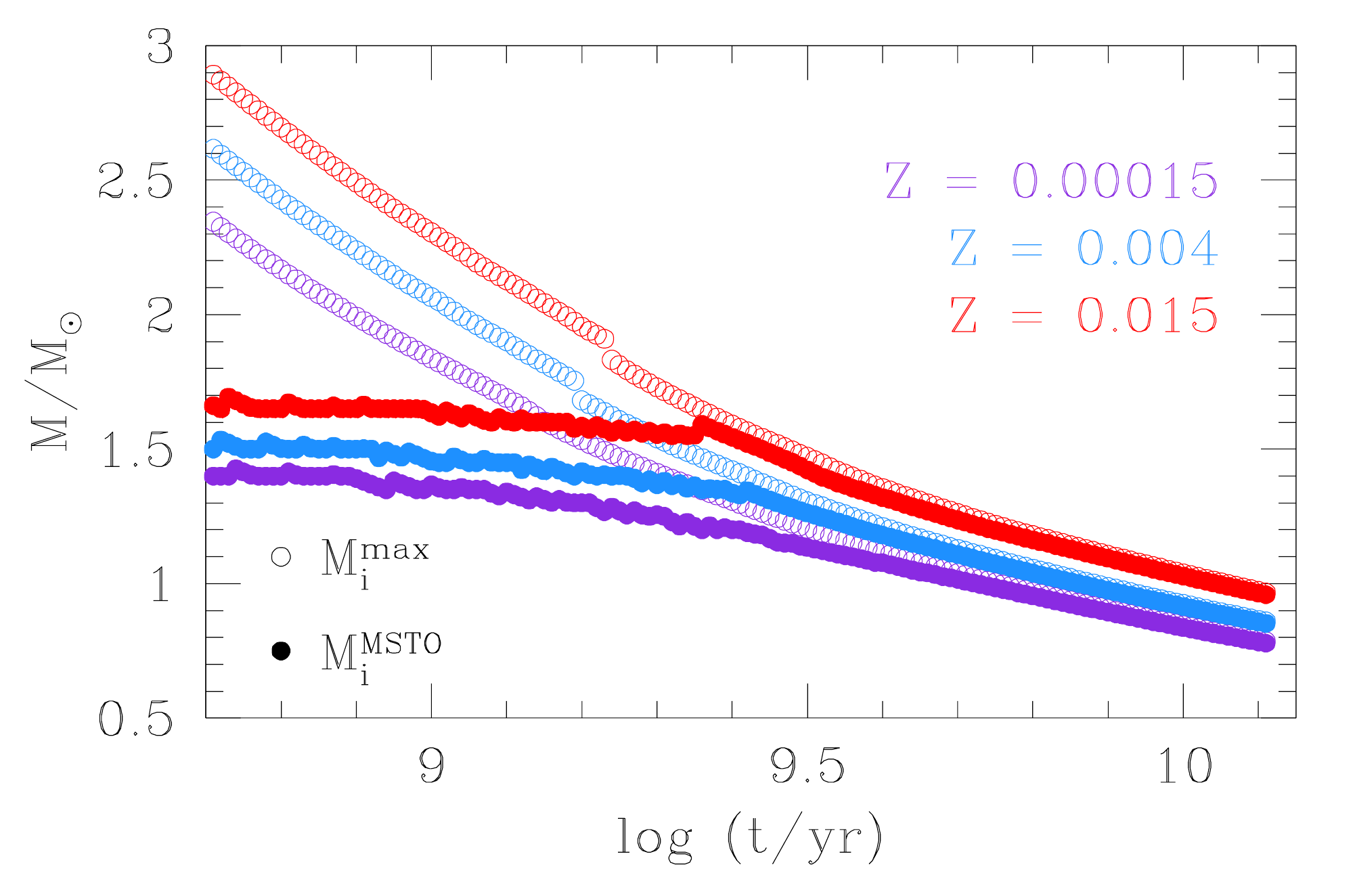}
\caption{Mass range of stars in post-main sequence phases as a function of age and metallicity. Data are from the PARSEC isochrones \citep{PARSEC}. Following \citet{Renzini86}, the initial mass of stars at the main sequence (MSTO) and the maximum initial mass of still-shining stars (M$^{\mathrm{max}}$) are shown. The number of stars in an observed sample of red giants depends on both the average red giant mass and the mass range on the RGB, introducing the possibility to bias sample statistics toward younger ages.}
\label{fig:fct}
\end{figure}

In a real observation, stars are selected for spectroscopy from some sample region in the CMD, frequently chosen to be within some well-defined colour and magnitude limits. In external galaxies, these are usually well above the main-sequence turnoff. This selection translates to the corresponding upper and lower mass limits, which are functions of age and metallicity.  An illustration of the impact of this choice is shown in Figure~\ref{fig:fct}, which shows the mass range of post-main sequence stars in the isochrones published by \citet{PARSEC} as a function of age and metallicity. For the purposes of computing relative numbers of stars, the initial mass of the star is shown, ignoring any mass loss during the stellar lifetime.

For ages younger than $\approx$2.5--3~Gyr, there is a steep increase in the mass difference between the main sequence turnoff and the most massive star still shining with decreasing age. This directly relates to the number of stars observable at any moment in time. This could partially cancel or even outweigh the tendency of IMF-weighting to produce more stars of low mass than high mass, which favours older stars in observed samples. At a given age, higher metallicity stars have a higher mass, which would tend to suppress their numbers through the IMF, but also a larger mass range, which would tend to increase their representation in a population.  In this way, an unbiased selection of stars from an observed CMD can produce an MDF which is skewed compared to the true distribution of heavy element abundances in the galaxy. 

The competing effects of age and metallicity on the mean mass and mass range of RGB stars are not easily disentangled analytically, but are amenable to simulation using stellar models if the IMF, and observational selection effects can be investigated.  The magnitude and direction of the effects illustrated in Figure~\ref{fig:fct} will also depend on what fraction of the total post-main sequence mass range is sampled, i.e., on the colour and magnitude range chosen for the spectroscopic sample.

In this paper we examine the distribution of stars in synthetic colour-magnitude diagrams (CMDs) created from theoretical isochrones to derive corrections that can be applied to RGB metallicity distributions of resolved stellar populations.  This procedure allows us to calculate the size and direction of the biases introduced into the metallicity distribution as a function of age and metallicity. This is expected to help identify areas of the (age, metallicity) parameter space in which an apparent lack (or excess) of stars is likely to be due to IMF and/or stellar evolution effects.  We show how such cases can result in spurious conclusions about nucleosynthetic or gas-dynamic processes sometimes applied in chemical evolution models.

\subsection{The Simulations}

We synthesised colour-magnitude data from isochrones generated using the on-line resource CMD 2.7\footnote{\verb= http://stev.oapd.inaf.it/cgi-bin/cmd =, 18-02-2016}, using version 1.2S of PARSEC - the \textbf{Pa}dova and T\textbf{r}ieste \textbf{S}tellar \textbf{E}volution \textbf{C}ode \citep[see][and references therein]{PARSEC} with the default options on mass-loss and circumstellar dust.  
For conversion to the observer plane, we used the colour-magnitude system for the Hubble Space Telescope Advanced Camera for Surveys filter set, as this is characteristic of the best-quality photometry of most Local Group dwarf galaxies. For our simulations, we used the ACS I-band (F814W) and V-band (F555W) bandpasses. 

\begin{figure}
	\includegraphics[scale=0.8]{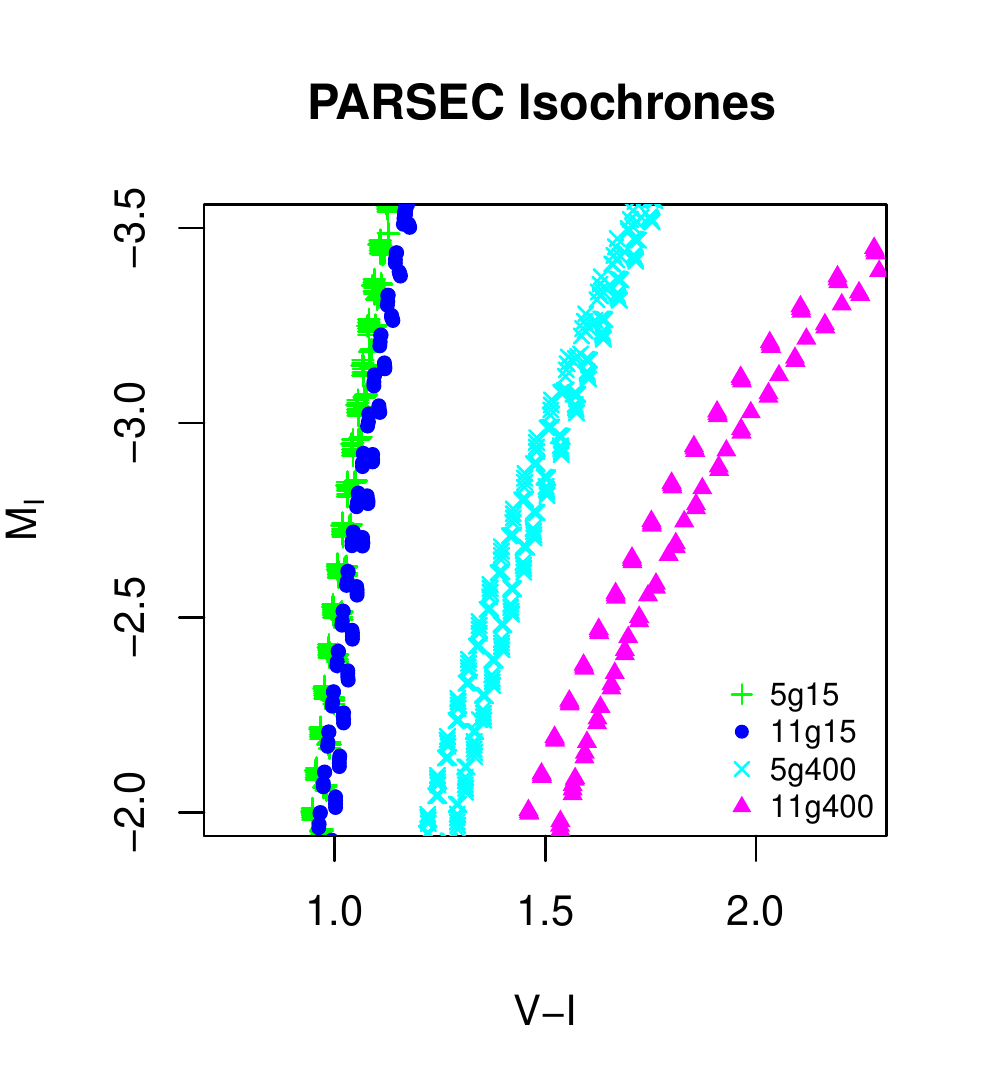}
	\caption{Comparison of four PARSEC isochrones, varying both age and metallicity, corresponding to four of our simulations (shown in Figure~\ref{fig:4CMDs}).  Simulation 5g15 has stellar ages of 5-6 Gyr with a metallicity of Z=0.00015; simulation 11g15 is 11-12 Gyr with Z=0.00015; simulation 5g400 is 5-6 Gyr with 0.004, and simulation 11g400 is 11-12 Gyr with 0.004.  See Sec.~\ref{sec:simulations} for details.}
	\label{fig:isochrones}
\end{figure}

The isochrones input into the simulations are shown in Figure~\ref{fig:isochrones} for four representative age-metallicity pairs and a typical range of absolute magnitudes.  The simulations were named in the following convention: the lower age bin limit serves as the prefix in Gyr and the metallicity ($\times10^{-5}$) is the suffix.  For example, for metallicity Z=0.004 and an age range of 3-4~Gyr, the simulation is called 3g400, while the age range of 0.5-1~Gyr for Z=0.004 is shortened to 0g400.

In the synthetic CMDs, each point in the original isochrones is assigned a weight according to the IMF; the weight is smoothed by interpolation between adjacent points to reduce granularity, and the weights are summed and binned to discretise the CMD. Whilst high-metallicity stars will tend to fall onto the red side of the RGB, they will be further spread in colour according to their age variation. On the blue side of the RGB, there will be a mix of young and old stars at low metallicity, plus the youngest stars of high metallicity.  Any observational sample of RGB stars drawn from a colour-magnitude window will contain a mixture of stars of whatever ages and metallicities are present, weighted by the mean mass and mass range of stars within the selection region.

We used the log-normal plus power law IMF from \citet{Chabrier01}; because the largest differences between this IMF and other commonly-used choices \citep[e.g.,][]{Salpeter55,Kroupa93} occur among the very low mass stars which do not evolve to the RGB over the ages simulated here, all of the relative RGB numbers should scale together, nearly independent of the choice of IMF.  We created synthetic CMDs in seven different age bins of 0.5-1, 1-2, 3-4, 5-6, 7-8, 9-10 and 11-12~Gyr, each with six different metallicities; Z = 0.00015, 0.0004, 0.0015, 0.004, 0.015 and 0.04 (equivalent to [M/H] = $-2, -1.6, -1, -0.6, 0, +0.4$, where [M/H] $\equiv$ log$_{10}$(Z/Z$_{\odot}$), with Z$_{\odot}$ = 0.0158), for a total of 42 simulations.  Four representative simulations are shown in Figure~\ref{fig:4CMDs}, while the full set are displayed in Appendix~\ref{app:42CMDs}.  The PARSEC isochrones shown in blue and cyan, represent the upper and lower age limits (respectively) of each simulation.  For example, simulation 1g15 has a lower limit of 1~Gyr and an upper limit of 2~Gyr.  The grey-scale points are the simulated stellar number densities in each colour-magnitude bin.  The simulated points scattering away from the isochrones are due to the presence of binaries; this most often manifests as a slight brightening relative to the isochrones, but can be more obvious when the secondary has not yet evolved off the main sequence and is therefore much bluer in colour than the primary.

To create our synthetic RGBs, we used isochrones interpolated to a very fine grid, spaced by 0.01 in log(age).  Each mass point along each isochrone was weighted by the integral of the \citet{Chabrier01} IMF over the interval centred on the mass point and extending halfway to the adjacent points.  Each point in the isochrones is thus translated to a probability density over a discrete interval.  Within a desired age interval, the isochrones are summed to form the total predicted number of stars in each CMD bin.  The weights are scaled to reproduce a constant star formation rate of 1 M$_{\odot}$/yr over the 1~Gyr span, integrated over all possible stellar masses.

As with previous works using CMDs to determine metallicity \citep[see][]{Casagrande11,Ordonez15}, we excluded Horizontal Branch (HB) stars from the sample, along with Blue Loop (BL), Red Clump (RC), AGB and post-AGB (PAGB) stars, where applicable.  These stars were omitted based on the CMD morphology of PARSEC isochrones via visual inspection and applied polynomial and logarithm functions (see Figure~\ref{fig:4CMDs} and Appendix~\ref{app:42CMDs}).  The HB and so the colour-cutoff is simply determined to be the point at which the colour-magnitude slope ($dI/d(V-I)$) of the locus of points in the simulation approaches zero.

The aim was to include all red giants in the sample, even at the expense of including some AGB and HB stars.  We can easily exclude the younger AGB stars, while older AGB stars tend to overlap the younger RGB, especially at lower metallicities (see top left panel, 1g15).  In older stars at higher metallicity (e.g., 11g4000, bottom right panel), it is much easier to exclude the AGB during sample selection.  Note that the differing degrees of AGB-overlap as a function of age and metallicity may introduce some inconsistency in our corrections.  However, since there are relatively few populations that contain both old, metal-rich and young, metal-poor stars, we believe this will not significantly affect the results.

\begin{figure*}
\includegraphics[scale=0.9]{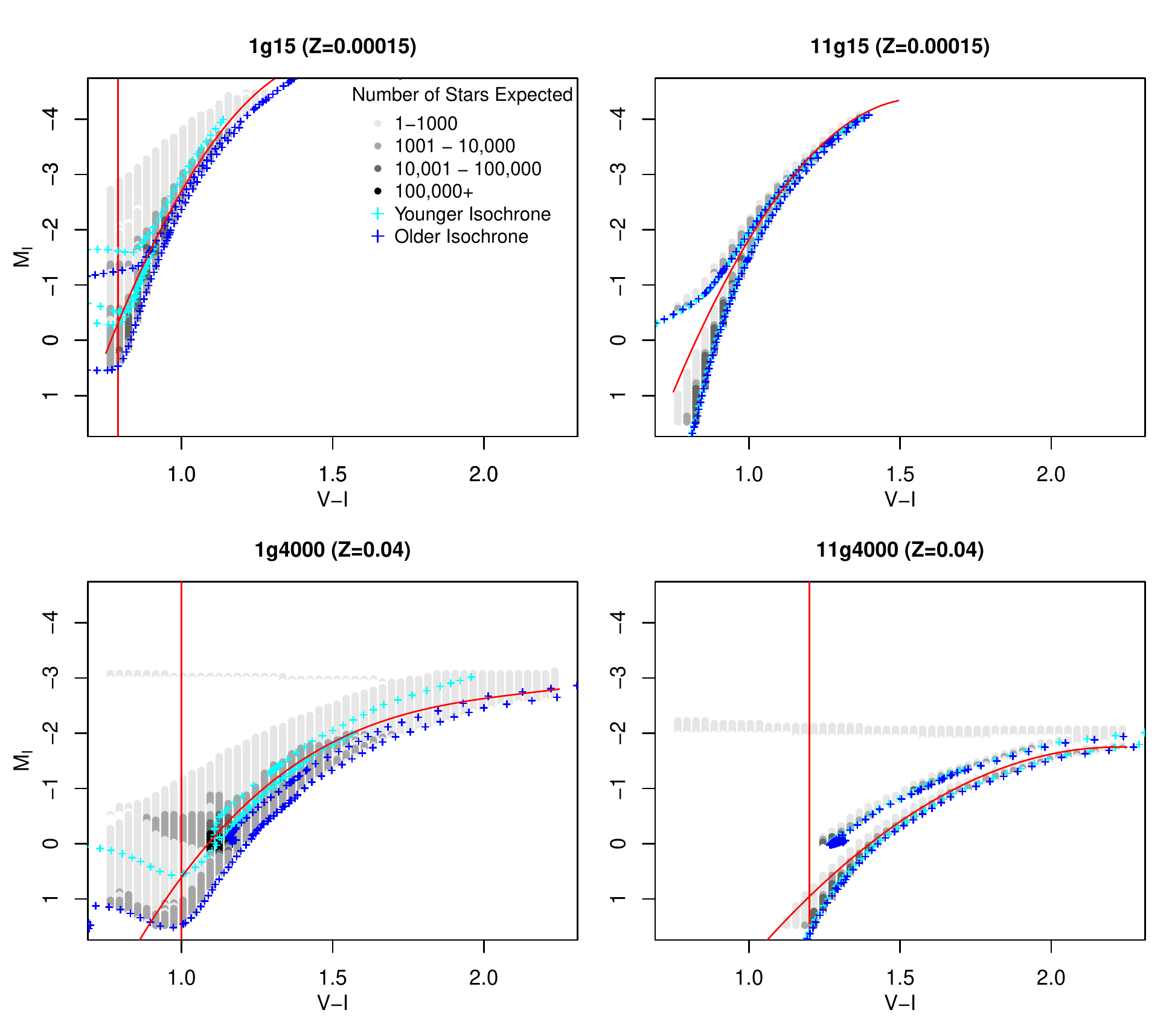}
\caption{Select four simulated CMDs, with varying age and metallicity.  The full 42 simulations are shown in Appendix~\ref{app:42CMDs}.  The legend shown in the top left plot applies to all four CMDs.  The red vertical line in each plot represents the individual BL and HB colour-cutoff determined for each simulation, where possible.  The red parabola is the cutoff between AGB and RGB stars.  The older and younger isochrones represent the upper and lower age limits of the simulation; in these cases, spanning a range of 1~Gyr.  See Sec.~\ref{sec:simulations} for details.} 
\label{fig:4CMDs}
\end{figure*}

Each simulation shows the calculated number and distribution of stars produced along the RGB if the SFR is a constant 1\msol per year for 1~Gyr.  The probability density distributions from the simulations are resampled to a regular grid of 6000 colour-magnitude bins for consistency across age and metallicity, with a magnitude range of $-4.5 \leq M_I \leq 1.5$ (bin width 0.05) and colour range of $0.75 \leq V-I \leq 2.25$ (bin width 0.03).  The mass range across each bin ($\Delta M_{ini}$) on the RGB is on the order of $\sim10^{-3}$\msol, depending on the age of the stars.”

These simulations do not include any synthetic broadening to account for observational errors, which is reasonable for the commonly encountered case in which the photometric limit of the data is 2 or more magnitudes fainter than the HB.  This allows us to explore the effects of observational selection on the recovered MDF in a general way, independent of the parameters of any single project.  In particular, we explore sampling effects in two limiting cases: a) For a given age, the apparent MDF that results from a uniform distribution of metallicities; b) For a single metallicity, the age distribution of observed red giants resulting from a uniform distribution in age. The first case is analogous to a star cluster or burst population that could have multiple metallicities present \citep[for example, the ultra-faint dwarfs investigated by][]{Brown14}, while the second corresponds to a constant SFR with a flat age-metallicity relation.

\section{Results}
\label{sec:results}

\begin{table}
	\centering
	\begin{tabular}{l|cccc}
		& \textbf{5g15} & \textbf{11g15} & \textbf{5g400} & \textbf{11g400} \\
		\hline
		\textbf{$\bar{\mathrm{M}}$} &  0.9826 & 0.7692 & 1.059 & 0.8831\\
		\textbf{$\Delta$M} & 0.0793 & 0.0710 & 0.1095 & 0.1020\\
		\textbf{M$_{int}$} & 190.6 & 83.07 & 264.8 & 82.13\\
	\end{tabular}
	\caption{Mean stellar mass ($\bar{\mathrm{M}}$), change in stellar mass ($\Delta$M), and the total mass integrated along the isochrone (M$_{int}$) is calculated for the select isochrones shown in Fig.~\ref{fig:isochrones} (ages 5-6 Gyr and 11-12 Gyr,  with Z=0.00015 and 0.004, corresponding to our simulations 5g15, 11g15, 5g400, 11g400 respectively).  Masses are given in units of \msol and are derived from the colour-magnitude range $0.75 \leq V-I \leq 2.25$ and $-3.5 \leq M_I \leq -2 $.  }
		\label{tab:isochrones}
\end{table}

Table~\ref{tab:isochrones} shows the comparison of stellar masses along the four isochrones shown in Figure~\ref{fig:isochrones}.  Values for the mean mass, the change in mass between bright and faint magnitude limits ($-3.5 \leq M_I \leq -2 $), and the total mass integrated along each isochrone for that range were calculated.  The stellar mass values change with both age and metallicity, but to different degrees, demonstrating the complicated degeneracy that exists between the two.  Figure~\ref{fig:isochrones} shows that metallicity has a greater effect on the shape of the isochrones than age, and at low metallicity the age effect diminishes. These masses determine the number of RGB stars within the colour-magnitude selection window for each of our 10$^9$~M$_{\odot}$ simulations.

We calculated the bias factors for correction of metallicity distribution functions measured from RGB stars by comparing the number of stars populating the spectroscopically-sampled areas of the CMD to the total number of stars formed in each simulation. The analysis was performed separately over two magnitude ranges $-4.5 \leq M_I <-2.5 $ and $M_I \geq -2.5$ in order to investigate trends in bright and faint stars separately.  The results can be expressed as a single number for each age and metallicity, representing the maximum fraction of all stars formed that could be spectroscopically measured in the given CMD selection region.

\begin{figure*}
\includegraphics[scale=0.8]{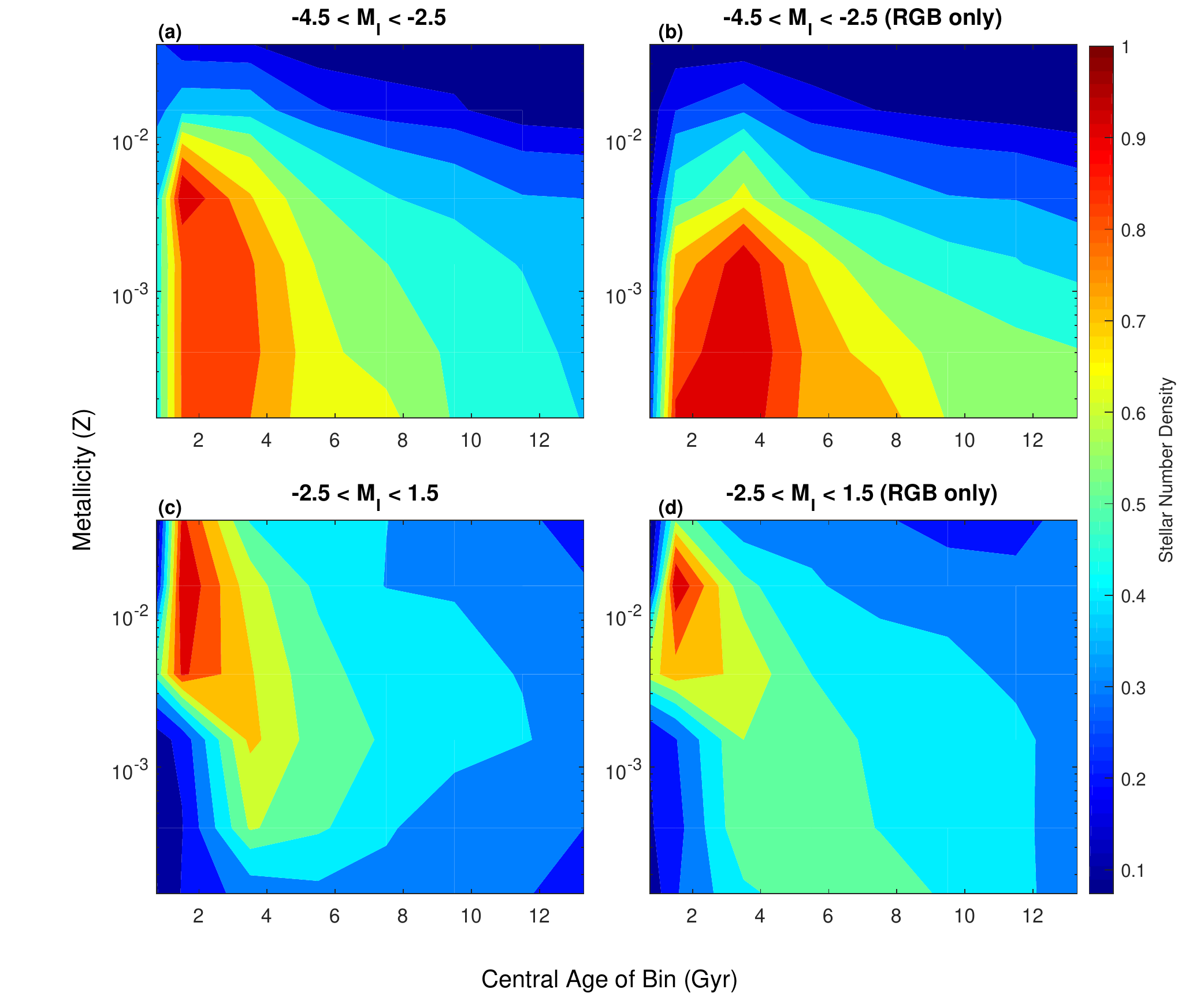}
\caption{The age-metallicity distribution of the 42 simulated RGBs over two magnitude (M$_I$) ranges.  Subplots $(b)$ and $(d)$ have the PAGB, AGB, BL, HB and RC stars removed from the sample (see Sec.~\ref{sec:simulations} for details).  The colour-scale background represents the predicted relative density of stars based on a constant SFR and flat AMR.  The values have been normalised for ease of comparison.}
\label{fig:contours}
\end{figure*}

We find a complex relationship between age, metallicity, and stellar density in our synthetic CMDs, differing greatly between the two magnitude ranges.  Figure~\ref{fig:contours} shows four contour plots: the top  images (subplots $(a)$ and $(b)$) show the brighter magnitude range, while the bottom (subplots $(c)$ and $(d)$) show the fainter range, cut off above the HB.  Panels $(b)$ and $(d)$ account for only the RGB stars, while panels $(a)$ and $(c)$ only exclude stars based on a minimum colour, and so include PAGB, AGB, HB, BL and RC stars.

The main result is that in a mixed population of stars, a sample of stars drawn from the RGB will be biased towards younger ages, making it less likely to detect older populations. While it could be na\"{\i}vely assumed that old stars are over-represented because of the IMF effect, here we show that in fact they are under-represented.  As demonstrated above and shown in Figure~\ref{fig:fct}, the number of stars observed is proportional to the integral of the luminosity from one magnitude level to another.  On the RGB this directly translates to an integral of the IMF from one mass to another. At younger ages this mass interval is larger than at old ages.

At a given age, generally low-metallicity stars are more likely to be observed than high-metallicity stars, but the size of the effect is not as dramatic as the variations with age.  While we anticipated that including fainter stars might improve the sampling by reducing the biases, this does not seem to be the case based on the comparison between the upper and lower panels of Figure~\ref{fig:contours}.  Panel $(c)$ in the Figure, shows that for fainter stars, the metallicity dependence is reversed.  Another factor is the larger separation of the RGB and the AGB at higher metallicities - the AGB of the older stars in the age range overlaps with the RGB of the younger stars, making removing the older AGB stars using morphology impossible.  This also increases the relative number of stars in this region of the CMD.

Singling out the RGB phase in the analysis gives very different age-metallicity distributions compared with the case when all the surrounding stellar phases (PAGB, AGB, BL, HB, and RC) are included.  This is shown in panels $(b)$ and $(d)$ of Figure~\ref{fig:contours}.  This highlights the benefit of clean separation between stellar sequences with high-quality photometry, where the star formation history (SFH) allows it.  A more detailed analysis of the relations between the age and metallicity for our synthetic populations is provided in Section~\ref{sec:dist} along with age and metallicity distributions (see Figures~\ref{fig:dist_age},~\ref{fig:dist_met}).

\subsection{Age and Metallicity Distributions}
\label{sec:dist}

\begin{figure*}
\includegraphics[scale=0.8]{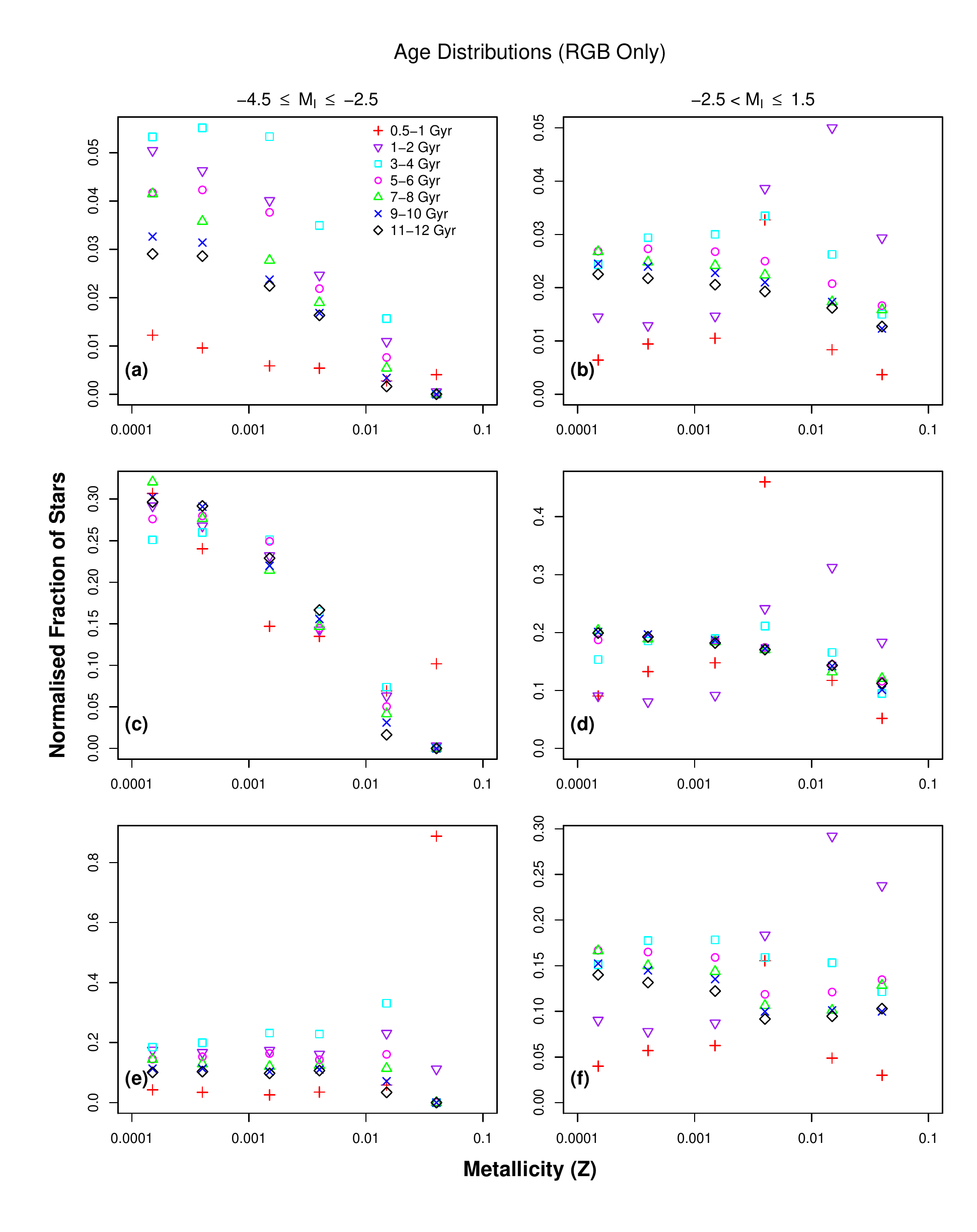}
\caption{Age distributions of red giants shown for both magnitude ranges.  The three rows of plots are normalised to different totals (see Sec.~\ref{sec:dist}), with panels $(a)$ and $(b)$ normalised to the total population, panels $(c)$ and $(d)$ normalised to each age population, while panels $(e)$ and $(f)$ are normalised to each metallicity population.  The legend shown in panel $(a)$ applies to all six plots.  These data are also presented in the contour plots in Fig.~\ref{fig:contours} (panels $(b)$ and $(d)$) and the metallicity distributions in Fig.~\ref{fig:dist_met} for comparison. Please note the different scales for the stellar number densities.}
\label{fig:dist_age}
\end{figure*}

\begin{figure*}
\includegraphics[scale=0.8]{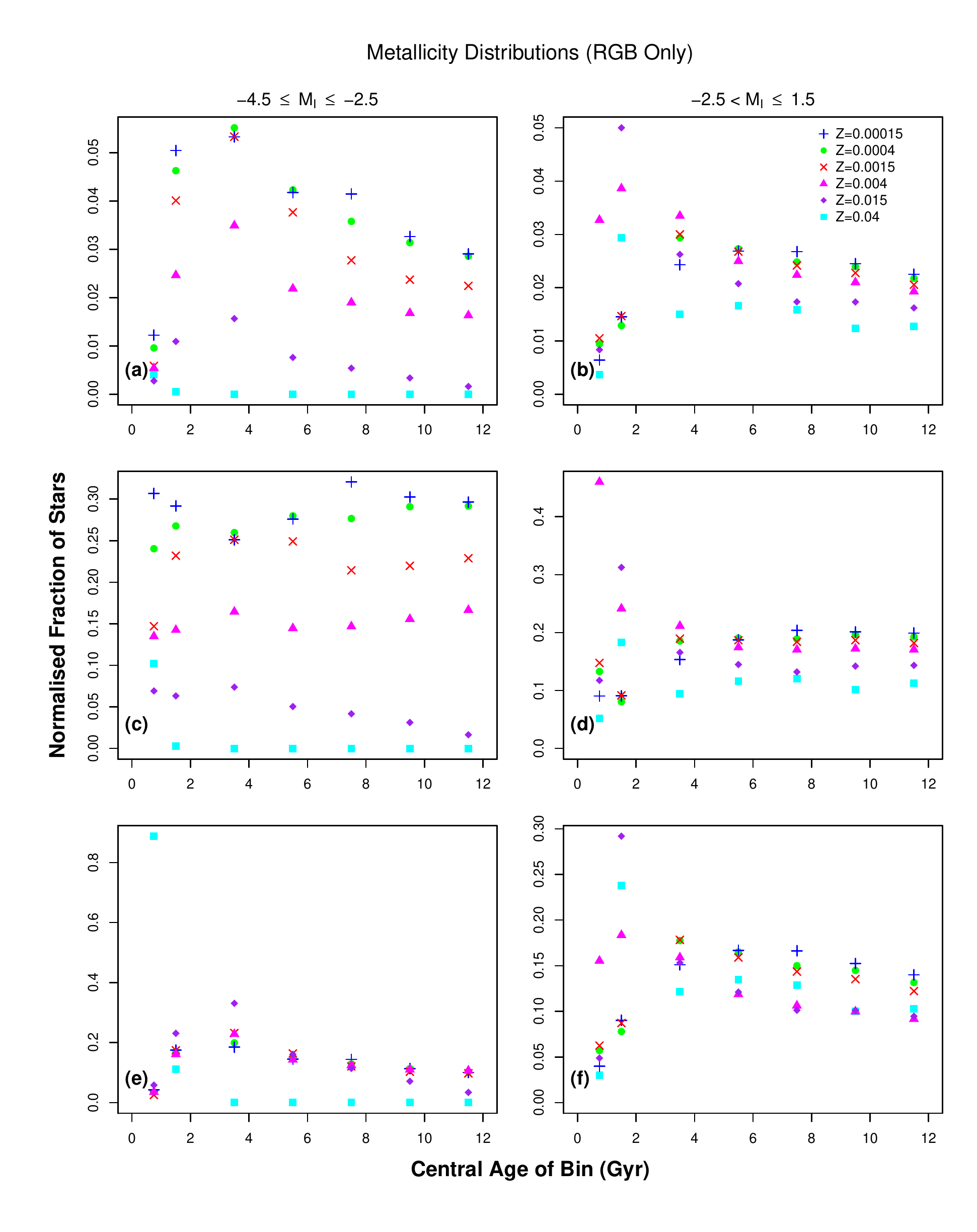}
\caption{Metallicity distributions of red giants shown for both magnitude ranges.  The three rows of plots are normalised to different totals (see Sec.~\ref{sec:dist}), with panels $(a)$ and $(b)$ normalised to the total population, panels $(c)$ and $(d)$ normalised to each age population, while panels $(e)$ and $(f)$ are normalised to each metallicity population.  The legend shown in panel $(b)$ applies to all six plots.  These data are also presented in the contour plots in Fig.~\ref{fig:contours} (panels $(b)$ and $(d)$) and the age distributions in Fig.~\ref{fig:dist_age} for comparison. Please note the different scales for the stellar number densities. }
\label{fig:dist_met}
\end{figure*}

In order to thoroughly investigate the behaviour within the contour plots (Figure~\ref{fig:contours}), we include here both metallicity and age distributions, normalised in three different ways, shown in Figures~\ref{fig:dist_age},~and~\ref{fig:dist_met}.  The first method was simply to normalise over the total stellar population across all 42 simulations, for each magnitude range.  The other methods were to normalise over either an entire metallicity or age population for each magnitude range.  For example, the simulation 11g15 was normalised over the total number of stars with ages between 11 and 12 Gyr in panels $(c)$ and $(d)$, and normalised with respect to the total number of stars with metallicities of Z=0.00015 in panels $(e)$ and $(f)$. 

Figures~\ref{fig:dist_age}~and~\ref{fig:dist_met} respectively show the two limiting cases we sought to investigate here: a) for a given age, the apparent MDF that results from a uniform distribution of metallicities; b) for a single metallicity, the age distribution of observed red giants resulting from a uniform distribution in age.  Each stellar age group shown in Figure~\ref{fig:dist_age} represents a population of stars formed in the same event (e.g. a star cluster), and their spread across a range of metallicities.  Figure~\ref{fig:dist_met} shows the observable numbers of red giants under the conditions of a flat AMR and constant SFR.

Figure~\ref{fig:dist_age} shows the age distributions for each normalisation method, over magnitude ranges of $-4.5 \leq M_I <-2.5 $ and $M_I \geq -2.5$ separately.  Shown clearly in these Figures, stars aged between 0 and 4 Gyr are the most unpredictable and often follow different trends to the older stars.  Each age sub-population follow similar trends at brighter magnitudes, whereas over the fainter interval, the metallicity relations in each age range behave very differently.  The artificial exclusions of the AGB, HB, BL and RC stars in our CMDs, which occur mainly at fainter magnitudes, could be partly responsible for panels $(b)$, $(d)$, and $(f)$ in Figure~\ref{fig:dist_age} being less smooth than their bright counterparts.  These exclusions were implemented manually for the most part and could introduce a small degree of inaccuracy into the data.  However, it could not account for the significantly different AMRs between the two magnitude selection ranges, as evident by the fact that the metallicity distributions show the opposite trend (Figure~\ref{fig:dist_met}).  This is further evidence that observations need to be carefully planned to avoid biases introduced by not properly considering the inclusion or exclusion of the fainter stars.  Figure~\ref{fig:dist_met} shows the metallicity distributions in a similar manner.  The plots in these Figures make obvious that the high-magnitude behaves very differently to the low-magnitude population, so careful modelling is required to interpret any observations.

The obvious outliers in panel $(e)$ of both figures represents the same data point.  It demonstrates that the vast majority of metal-rich red giants (here Z=0.4) are very young (0.5-1~Gyr).

\section{Discussion}
\label{sec:discussion}

We have investigated the relationship between the observational sample selection of red giant stars and the true metallicity distribution by simulating 42 CMDs with varying ages and metallicities.  Comparing the number of stars populating each region of the CMD allowed for the calculation of approximate correction factors to implement in current models.  While the need to account for variable stellar lifetimes was appreciated very early on in the integrated light study of galaxy spectra \citep[e.g.][]{Renzini86} and is easily taken into account by the procedures needed to fit whole-of-population line indices \citep[see for example][]{Trager05}, the issues in drawing inferences from samples of individually selected giants have been explored much less thoroughly.

\subsection{Bias Corrections Applied to Real Data}
\label{sec:applications}

The synthetic RGBs constructed here assume a constant SFH and a flat AMR, thus the number density contours in Figure~\ref{fig:contours} cannot be applied directly to a real stellar system without first accounting for the SFH and chemical evolution.  \citet{Cole09} and \citet{Dolphin16} have advocated simultaneous modelling of the SFH from deep-photometry and spectroscopy of a subset of bright stars but in many cases this is not practical.  The age-metallicity contours in Figure~\ref{fig:contours} were constructed using a constant SFH and a flat AMR, which is equivalent to assuming no chemical enrichment over time, and no gaps or spikes in the star formation rate.  To apply a correction, the number of stars in each age band should be scaled by the SFR.  Some degree of knowledge about the SFH is required in order to apply these corrections, which is a limitation inherent to the use of chemical evolution probes with finite lifetimes.

\subsubsection{Dwarf Irregular Galaxies}
\label{sec:kirby}

As an example of how to apply and use our corrections, \citet{Kirby17} calculated scaling factors for metallicity distributions of Leo A and Aquarius, based on the SFH and AMR from \citet{Cole14} and the simulations presented here.  Although the correction did not have much effect on the Aquarius metallicity distribution, the mean metallicity of Leo A was calculated to be 0.07 dex lower than the observed value.  The overall shape of the Leo A metallicity distribution was also affected.  Changes in the mean metallicity and the distribution shape may lead to a different chemical evolution model being a better representation of the data.  In this case, the corrections to the metallicity distribution of Leo A resulted in the shape being slightly less peaked, meaning that the pre-enriched model was more favoured over the accretion model than from the analysis of the uncorrected data \citep[][see their Figure~10]{Kirby17}.

\subsubsection{Large Magellanic Cloud Bar}
\label{sec:LMC}

\citet{Cole05} presented spectroscopic metallicities for 373 red giants in the LMC bar, leading to a very well-defined observed MDF.  However, it is reasonable to suppose that the true MDF differs from this; we are now in a position to model the difference directly.

Because the SFH of this region of the LMC is well-constrained, we can use our synthetic RGBs to compare the predicted distribution of red giants to the observed distribution, as well as to provide a corrected MDF for chemical evolution modelling (CEM).

\begin{figure*}
	\centering
	\includegraphics[scale=0.8]{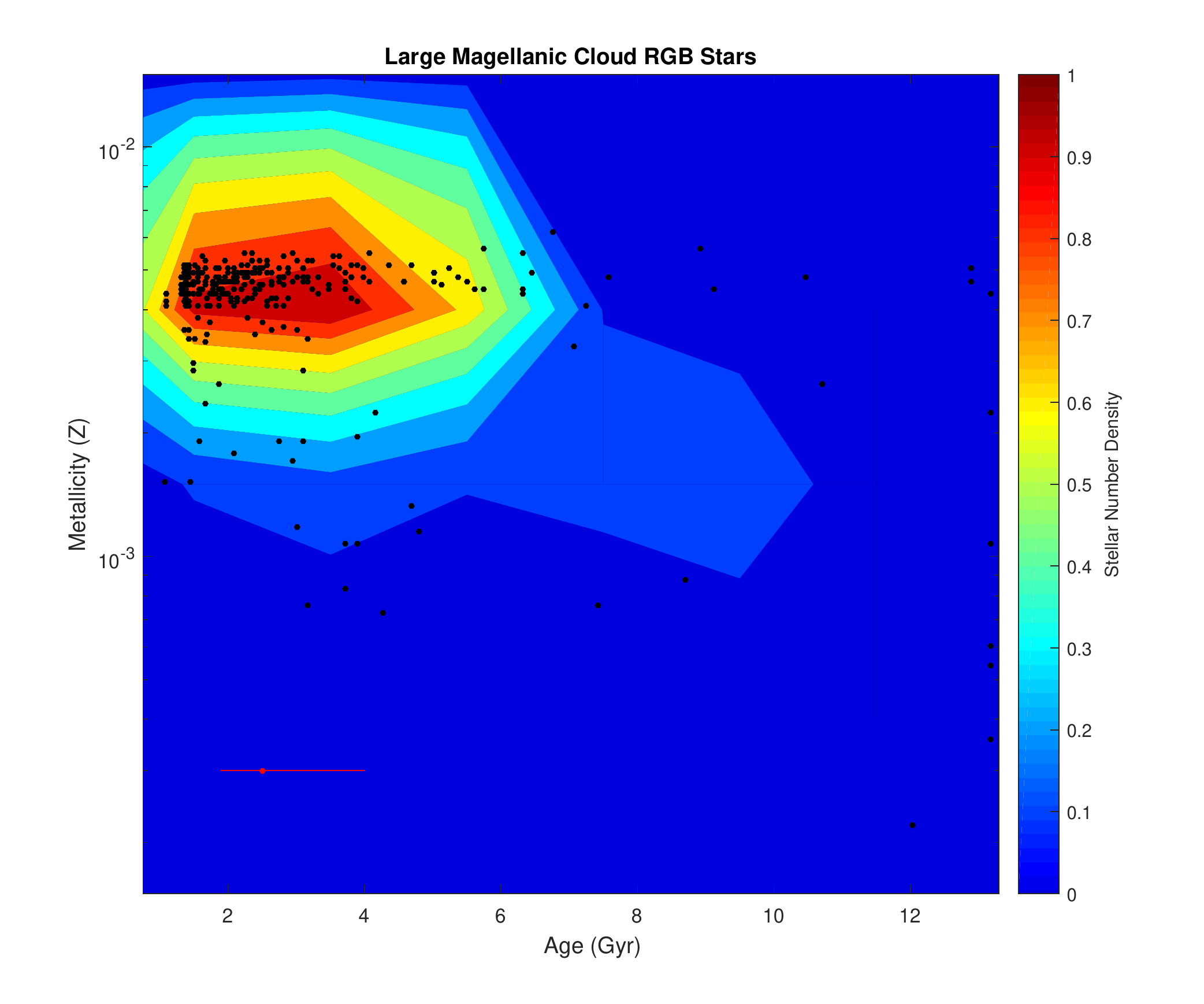}
	\caption{Age-metallicity distribution from simulated RGBs, for LMC bar distance reddening, AMR, and SFH.  See Sec.~\ref{sec:LMC} for details.  The black points are the LMC bar RGB stars taken from \citet{Cole05} and the red point is a representative error bar for the uncertainty in the stellar age.  The colour-scale background represents the relative density of stars in our simulations, scaled by the SFH of \citet{Cole09}.  Although the SFH is greatly extended, most RGB stars are expected to be 1.5-4 Gyr old.}
	\label{fig:Cole05}
\end{figure*}

Only RGB-phase stars were included in the simulated data.  We adopted a distance modulus of 18.50, also used by \citet{Cole05}.  Hence, the RGB stars in the LMC bar were estimated to be $-3 \leq M_I \leq -2 $, which sits comfortably within our simulated range.  The 373 observed RGBs from \citet{Cole05} are shown overlaying the simulated metallicity distribution contours in Figure~\ref{fig:Cole05}.  We adjusted the high-end metallicities following \citet{vanderSwaelmen13} and re-derived the ages using the same method as \citet{Cole05}, but using the PARSEC isochrones \citep{PARSEC} for consistency.  

The SFH (which includes the modelled $\psi(t)$ and age metallicity relation, derived from broad-band photometry) was taken from \citet{Cole09}.  The predicted distribution of RGB stars for the LMC bar is in good agreement with the distribution of observed red giants.  At ages less than $\approx$4~Gyr, the metallicity distribution predicted from the CMD is broader than the observed MDF, because the SFH derived from broad-band colours does not strongly constrain [Fe/H].  The distribution of RGB ages predicted by our simulations scaled by the SFH is in good agreement with the distribution of ages predicted from the RGB colours and metallicities alone. This serves to emphasise that the observed RGB sample is biased towards young ages.  The median age of the observed sample is 1.9~Gyr while the median age of star formation is $\approx$5~Gyr.  The predicted metallicity distribution of LMC bar red giants suggests that stars more metal-poor than the peak are under-represented by $\approx$25\% (see Figure~\ref{fig:hist}).

\subsubsection{Chemical Evolution Models}

\citet{Cole05} did not fit chemical evolution models to their sample, but some work was done in this area by \citet{Carrera08}.  In order to overcome the age-metallicity degeneracy in RGB stars \citep[see][]{Worthey94}, \citet{Carrera08} combined spectroscopy with deep CMD photometry to calculate the ages for individual RGB stars in the LMC.  While this is a good approach in that the MDF alone is not used to determine the ages, the authors did directly use the number of RGB stars to determine the SFH, likely biasing their results against old, metal-poor stars.

We tested the impact of our correction to the MDF derived above by fitting simple chemical evolution models to the corrected LMC data following \citet{Carrera08} \citep[see also][for original derivations]{Tinsley80,Peimbert94}, using the instantaneous recycling approximation.  The metallicity of the system as a function of time is determined via

\begin{equation}
\mu\frac{dZ}{d\mu} = \frac{y(1-R)\psi+(Z_f-Z)f_I}{-(1-R)\psi+(f_I-f_O)(1-\mu)}
\label{eq:ODE}
\end{equation}

\noindent where $\mu$ is the ratio of the gas mass to the total baryonic mass, $y$ is the yield, $R$ is the mass fraction returned to the system after each generation of stars, compared to the total mass of the stars in that generation, $Z_f$ is the metallicity of the accreted gas, and $f_I$ and $f_O$ are the accretion and outflow rates, respectively.  The simplest case is that of a pre-enriched closed box with no gas flowing in or out ($f_I=f_O=0$).  Integrating Equation~\ref{eq:ODE} to solve for $Z(t)$ gives

\begin{equation}
Z(t)=Z_i+y\mathrm{ln}\mu(t)^{-1}
\label{eq:CB}
\end{equation}

\noindent where $Z_i$ is the initial metallicity of the system.  For the case where gas is accreted but none is lost ($f_O=0$), we follow \citet{Carrera08}, defining the accretion rate to be $f_I=\alpha(1-R)\psi$, where $\alpha$ is a free parameter.  We make the not unreasonable assumption that accreted gas has the same metallicity as present initially in the system (that is, $Z_f=Z_i$).  In the case that $\alpha\neq1$, Equation~\ref{eq:ODE} becomes

\begin{equation}
Z(t)=Z_i+\frac{y}{\alpha}\left[1-\left(\frac{\mu(t)}{\alpha(\mu(t)-1)+1}\right)^{\alpha/(1-\alpha)}\right] .
\label{eq:Acc1}
\end{equation}

\noindent When $\alpha=1$, the accretion model can be expressed as

\begin{equation}
Z(t)=Z_i+y\left[1-e^{1-\mu(t)^{-1}}\right] .
\label{eq:Acc2}
\end{equation}

\noindent Lastly, we consider a leaky box scenario with no gas being accreted ($f_I=0$), where the rate of gas flowing out of the system is $f_O=\lambda(1-R)\psi$.  $\lambda$ is also a free parameter, and for the case when $\lambda\neq1$, Equation~\ref{eq:ODE} becomes 

\begin{equation}
Z(t)=Z_i+\frac{y}{\lambda+1}\mathrm{ln}\left[\frac{\lambda+1}{\mu(t)}-\lambda\right] .
\label{eq:LB}
\end{equation}

\begin{figure}
	\includegraphics[scale=0.7]{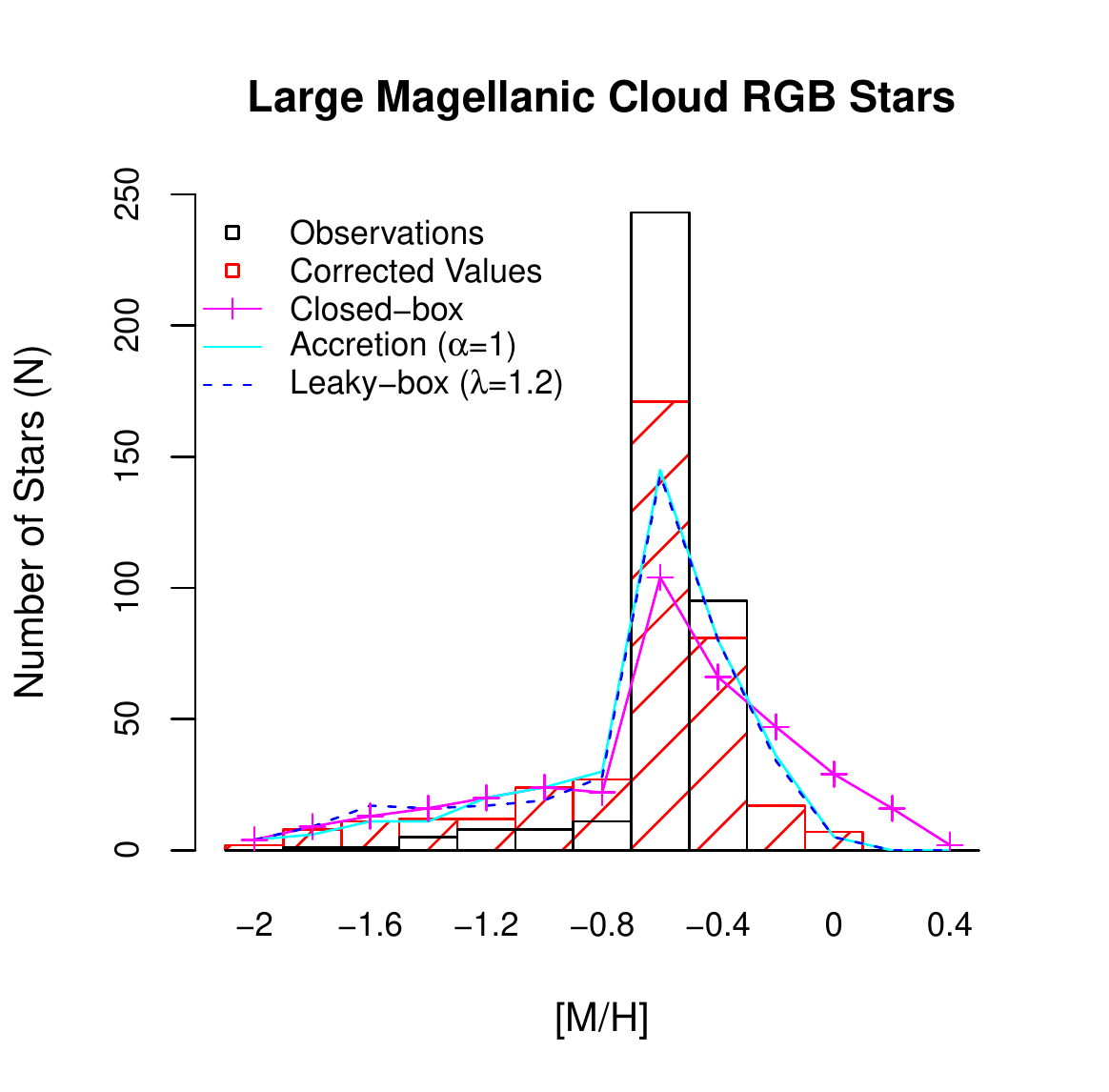}
	\caption{Metallicity distribution of the observed LMC bar red giants taken from \citet{Cole05} shown in black, compared with our corrected MDF shaded in red, along with CEMs adopted from those in \citet{Carrera08}.  The full AMRs are shown in Fig.~\ref{fig:CEMs}.}
	\label{fig:hist}
\end{figure}

\begin{figure}
	\includegraphics[scale=0.8]{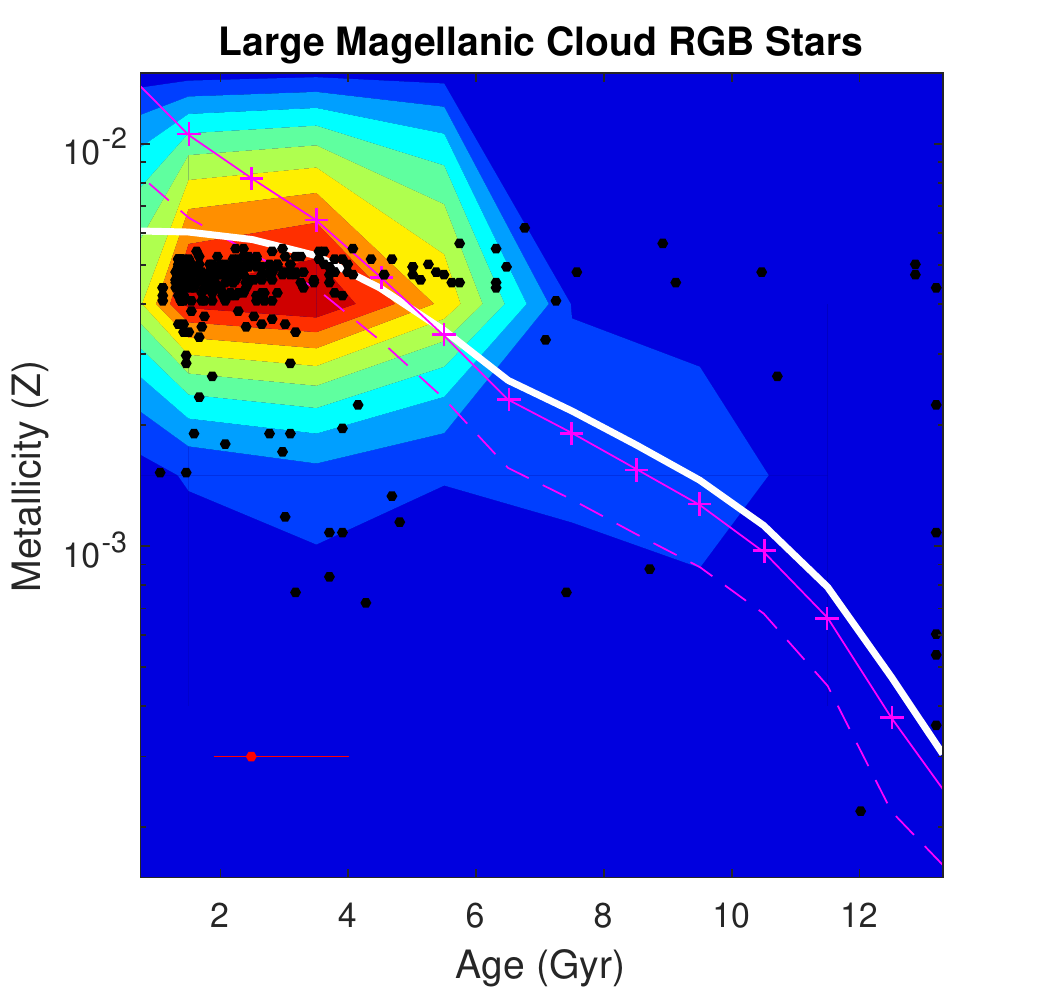}
	\caption{Same as in Fig.~\ref{fig:Cole05}, the black points are the observed RGB stars from the LMC bar, with the contours showing our corrected AMR.  The best-fit model is the accretion model using $\alpha = 1$ and is shown as the solid white line.  The magenta dashed line shows the leaky-box model with $\lambda = 1.2$, and the closed-box model is shown with the crossed magenta line.  The same CEMs are shown in Fig.~\ref{fig:hist}.}
	\label{fig:CEMs}
\end{figure}

Figure~\ref{fig:hist} shows the metallicity distribution of the observed LMC bar red giants \citep[taken from][]{Cole05} compared with our corrected MDF, along with the three simple CEMs; pre-enriched closed box, accretion, and leaky box.  We chose a yield of 0.006, in alignment with the current metallicity of the LMC.  For the accretion and leaky box models we chose values of $\alpha$ and $\lambda$ based on the goodness of fit when applied to the metallicity distribution.  Equations~\ref{eq:CB},~\ref{eq:Acc2},~\ref{eq:LB} were directly used to plot the AMR from each of the models over the LMC stellar distribution in Figure~\ref{fig:CEMs}.  We investigated the effect of including a SFR contribution from the LMC disc, but for reasonable values the changes to the AMRs were not significant.

The metallicity distribution is best represented by either an accretion or a leaky-box model when corrections for the sampling bias are taken into account (see Figure~\ref{fig:hist}).  The only differences between the two are in the low-metallicity tail.  \citet{Carrera08} proposes a model featuring gas moving both in and out of the system as the best representation, with free parameters of $\alpha = 1.2$ and $\lambda = 0.4-0.6$.  This is equivalent to a nett inflow with $\alpha = 0.6-0.8$, conditions that we found were not significantly different to the closed-box model when representing the corrected metallicity distribution.  Interestingly, the age-metallicity relation is not well-reproduced by any simple model, because of the large scatter in metallicity at intermediate ages.  Increasing the yield value used in the models \citep[to be closer to the value used by][]{Carrera08} better reproduces the peak metallicity in the distribution, but shifts the AMR too far to the young, metal-rich corner.  The SFH predicts that only a small number of RGB stars older than 6 Gyr should be observed, in agreement with observations, but the high metallicity of stars from 6--12~Gyr is not well-fit by the simple models. The cohort of metal-poor stars at ages 2--6~Gyr is also problematic for the chemical evolution models. It is likely that both more accurate age estimates for the RGB stars are required and more complex models need to be considered, but this is beyond the scope of this paper.

\subsection{Future Improvements}

In order to apply the corrections derived here the stellar age distribution needs to be roughly known; ideally this approach has to be applied to combined colour-magnitude diagrams and chemical abundance data.  While our corrections eliminate/alleviate some degree of the bias present in observed metallicity distributions, there is one main potential source for inaccuracy in our analysis: determining the HB and BL cutoff point and the separation of the RGB from AGB stars.  In other words, how to related the models to the observed sample.  As outlined in Section~\ref{sec:simulations}, the HB, RC, BL, AGB and PAGB phases were excluded from the sample, while keeping as many RGB stars included as possible.  Since the position on the CMD of the HB varied with both age and metallicity, it was not plausible to employ one magnitude limit for every simulation.  The colour-magnitude cutoffs are shown in the simulated CMDs in Appendix~\ref{app:42CMDs}, and visibly spread out at younger ages and higher metallicity, increasing the uncertainty in the analysis of these simulations in the dimmer magnitude range.  Whenever giant star samples are analysed to produce chemical evolution models, the stellar lifetime biases should be analysed using identical selection criteria to the data sample.

The equations derived to RGB stars from the simulations were determined through visual inspection of the separation between the RGB and AGB tracks, as well as the locations of any BL, RC and any PAGB stars, for each CMD.  To maintain consistency, the separation equations were all given as polynomial or logarithm functions.  The separation between the AGB and RGB was less pronounced at younger age ranges (0-1 and 1-2 Gyr), but at higher metallicities, the older stars on the AGB over-lapped the younger RGB stars for each age limit (see Appendix~\ref{app:42CMDs} for CMD plots of the simulations).  

\subsection{Comparison to Previous Work}

In a series of papers written by \citeauthor{Kirby11a} (\citeyear{Kirby11a,Kirby11b}), the authors derive a SFH using metallicities and $\alpha$-element abundances.  While \citet{Kirby11a} noted that their selection from a spatial subsection of the galaxies favoured young, metal-rich stars, they do not account for the fact that within this sample stellar lifetimes bias them further against old, metal poor stars.  A corrected MDF could make the conclusions in \citet{Kirby11c} regarding gas flows even stronger

\citet{Ross15} considered photometric rather than spectroscopic measures of the metallicity to examine the MDF of dwarf galaxies.  Traditional methods involving the colour of the RGB as a function of magnitude are subject to the age-metallicity degeneracy, but \citet{Ross15} neatly circumvent that problem by considering colour-colour plots, thus isolating metallicity as the measured parameter.  They demonstrate how their method produces observed MDFs that differ from those measured by \citet{Kirby13} based on spectra of individual giants.  While their method samples all giant stars that are present, it still does not account for the fact, demonstrated here, that their MDFs may still be biased by the over- or under-representation of stars of differing ages.  In principle their approach could be generalised by using synthetic colour-colour plots based on isochrones with a wide range of ages.

However, this is a sample selection correction based on the CMD distribution of stars and not on their variable lifetimes.  It is well known that metal-rich stars tend to be redder than metal-poor stars for a given stellar age.  For metallicities [M/H]$> -1$, the RGB tip no longer occurs at constant $I$ magnitude, so very metal-rich stars may be under-represented\citep{Reitzel02}.  Thus any selection based on colour could be removing a portion of either the more metal-rich or metal-poor populations.  \citet{Ho15} investigated some of the biases present in current RGB samples, applying corrections to their own MDF.  While the sample selection is necessary, it does not take the place of the stellar lifetime corrections derived here.

\section{Conclusions}
\label{sec:conclusions}

We examined the relationship between an apparent unbiased sample of red giant stars and the true metallicity distribution of the stellar population as a whole.  Synthetic RGBs based on PARSEC isochrones were used to produce the expected number distribution in terms of age and metallicity for a constant SFH and a flat AMR.  We present simple correction factors for over- and under-represented young to intermediate-age red giants at various metallicities.
  
In the case of galaxies where nearby all the stars are older than $\approx$10~Gyr \citep[e.g.][]{Brown14}, the mean metallicity and shape of the MDF will not be strongly affected, even where a broad ranges of metallicities is present.  Applying the corrections to Leo A in \citet{Kirby17} suggests that when these biases are taken into account, conclusions about the history of gas-flows and consumption can be significantly altered.  Where many intermediate age stars are present, the lifetime bias is expected to be more severe. 

We apply our trial corrections to the MDF and AMR of the LMC bar and show that the distribution of RGB stars based on the published SFH is in agreement with the observed distribution, which is biased towards the younger ages than the typical star.  The corrected metallicity distribution shows that metal-poor stars are being under-represented in the observed sample from \citet{Cole05}.  By updating the parameters used in \citet{Carrera08}, we find simple accretion and leaky-box models reproduce the shape and peak value of the metallicity distribution of our corrected stellar data, whereas applying the best-fit models from \citet{Carrera08} reproduced the overall shape well but the metallicity distribution was shifted towards the metal-rich end.

Ultimately the best results will be obtained by simultaneously modelling the CMD and red giant MDF to derive the SFH and AMR in a fully self-consistent way.  When this is not possible, great care must be taken when interpreting RGB metallicities using CEMs.

\section*{Acknowledgements}

The authors thank the anonymous referee for their comments that improved the clarity of this paper.  The authors are also thankful for discussions with Evan Kirby during the preparation of the manuscript that led to an improved paper.  AAC thanks Kim Venn for a provocative discussion about the initial mass function.  This research has made use of NASA's Astrophysics Data System Abstract Service.

\bibliography{References}

\appendix

\section{Simulated Colour-Magnitude Diagrams}
\label{app:42CMDs}

Our 42 synthetic CMDs of red giant stars (see Sec.~\ref{sec:simulations} for details).  The legend shown in the top left plot applies to every plot.  The red horizontal line in each plot represents the individual BL and HB colour-cutoff determined for each simulation, where possible.  The red parabola is the cutoff between AGB and RGB stars.  The younger and older isochrones represent the younger and older age limits.  

\begin{figure*}
	\centering
	\includegraphics[scale=0.9]{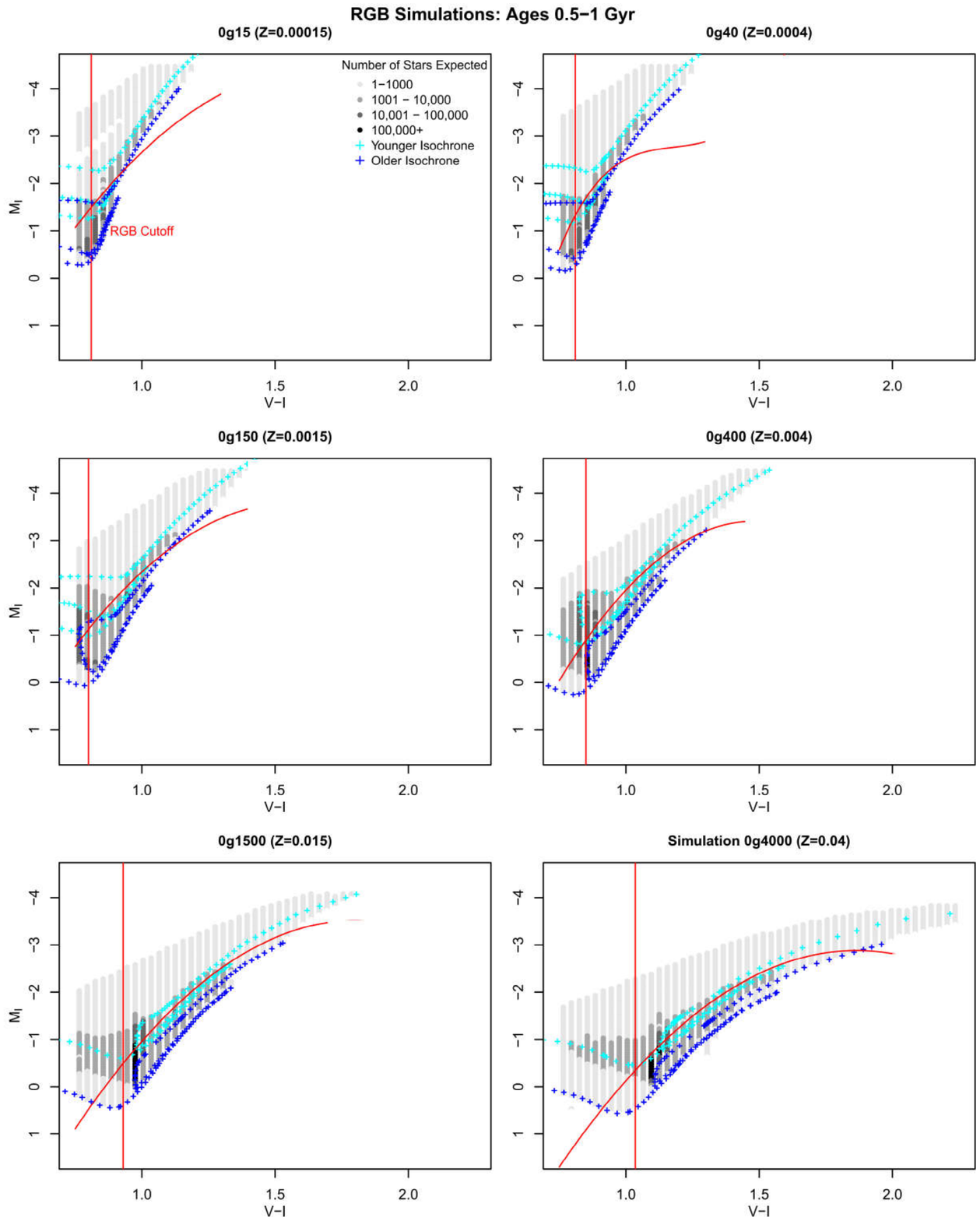}
	\caption{}
	\label{fig:app1}
\end{figure*}

\begin{figure*}
	\centering
	\includegraphics[scale=0.9]{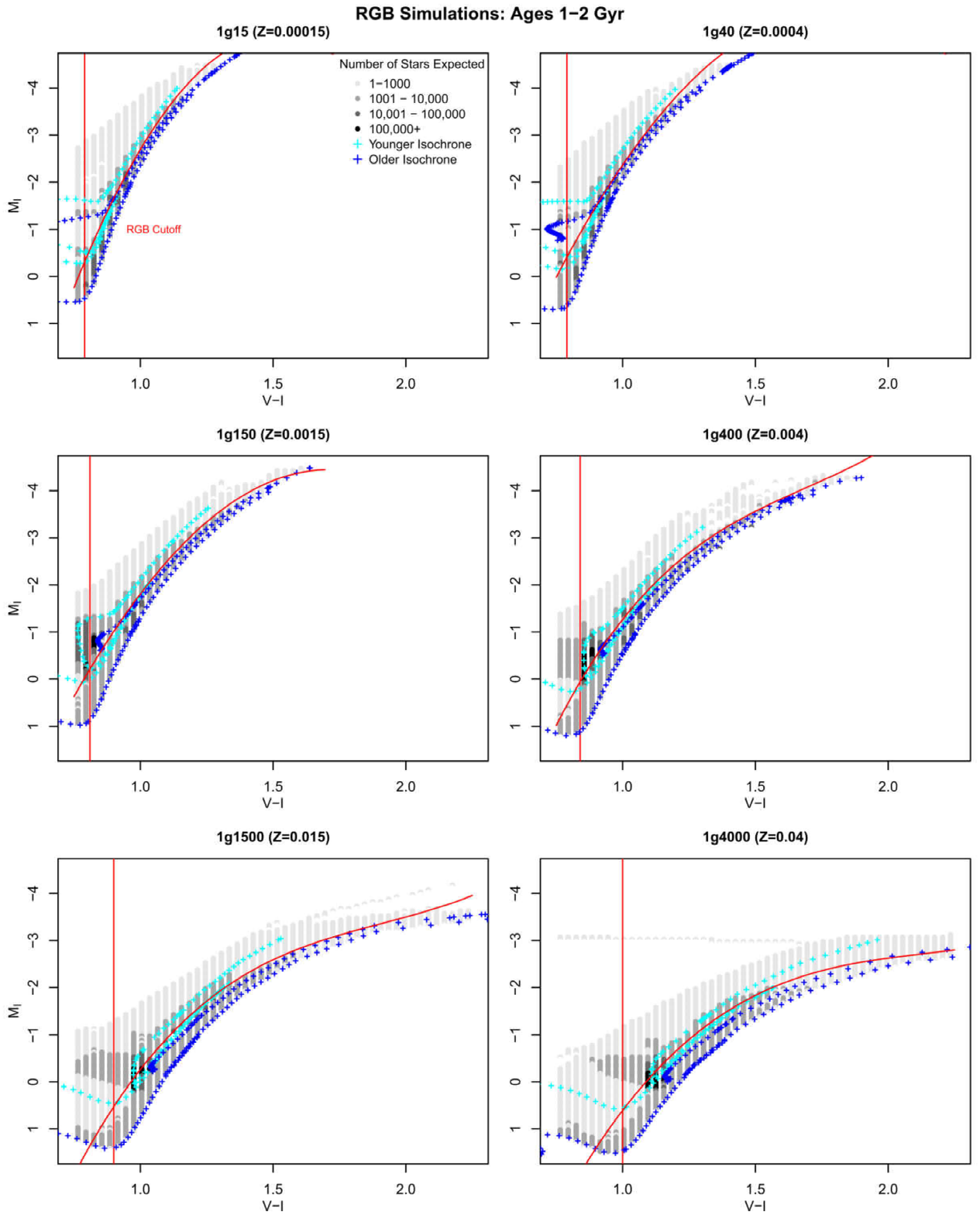}
	\caption{}
	\label{fig:app2}
\end{figure*}

\begin{figure*}
	\centering
	\includegraphics[scale=0.9]{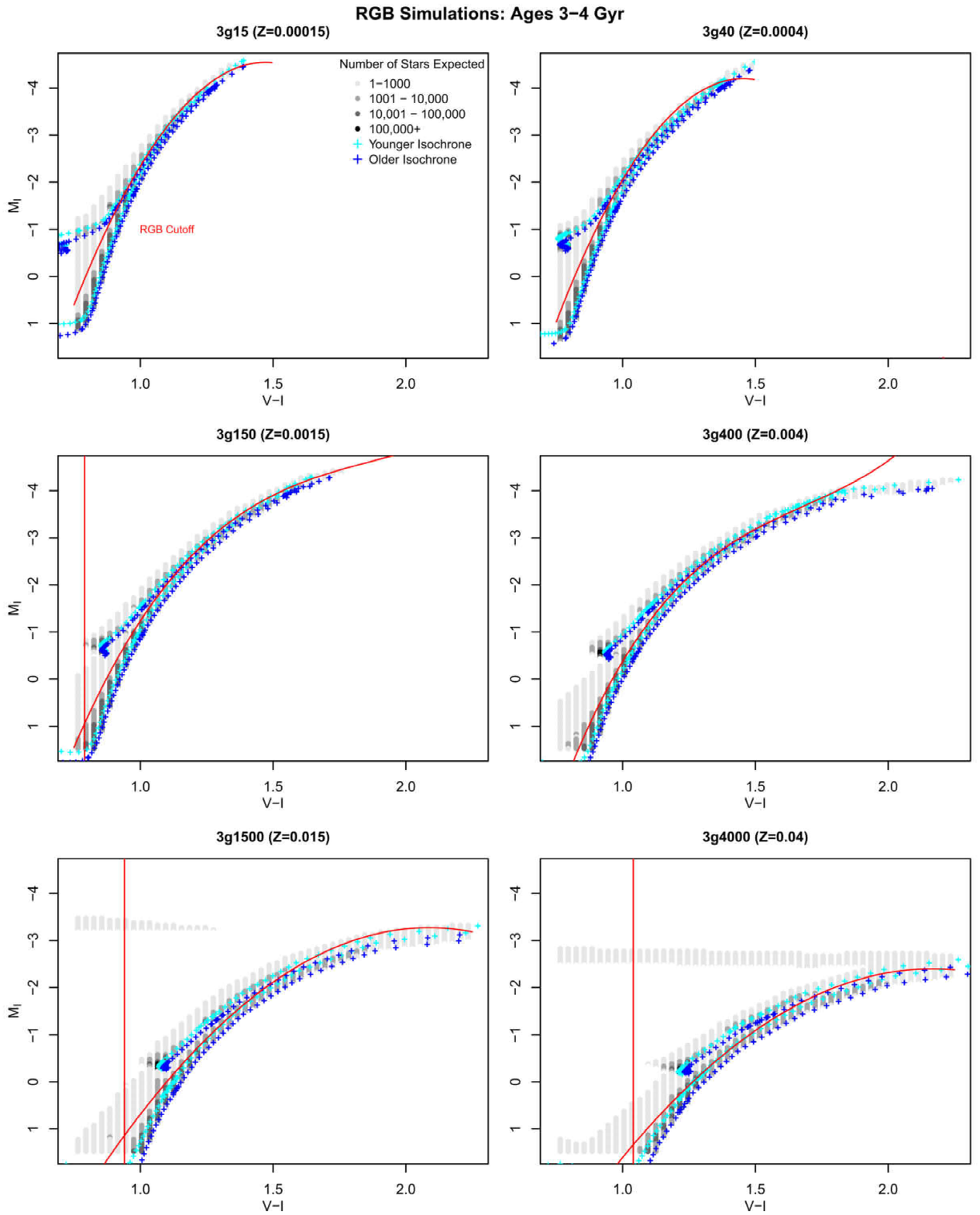}
	\caption{s}
	\label{fig:app3}
\end{figure*}

\begin{figure*}
	\centering
	\includegraphics[scale=0.9]{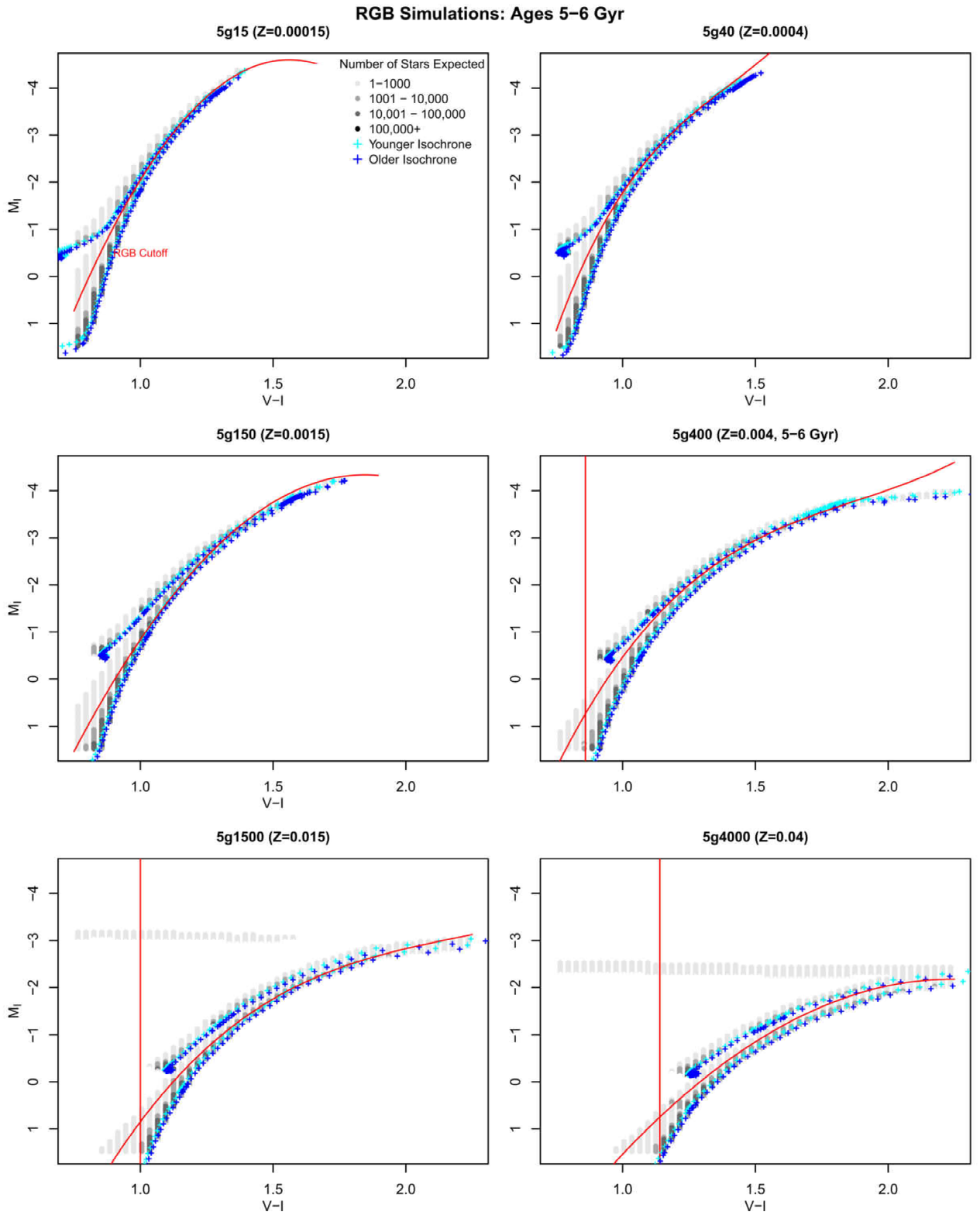}
	\caption{}
	\label{fig:app4}
\end{figure*}

\begin{figure*}
	\centering
	\includegraphics[scale=0.9]{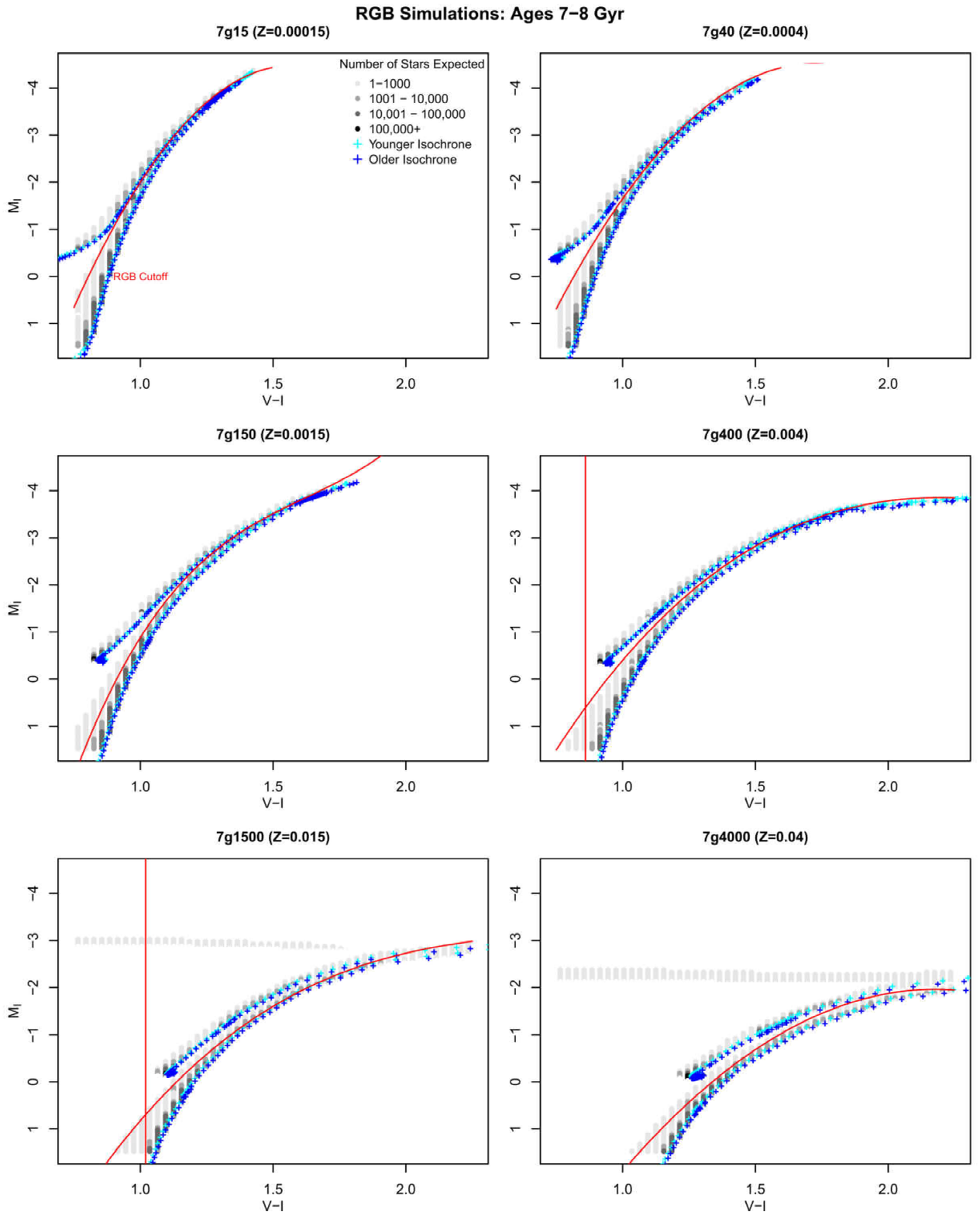}
	\caption{}
	\label{fig:app5}
\end{figure*}

\begin{figure*}
	\centering
	\includegraphics[scale=0.9]{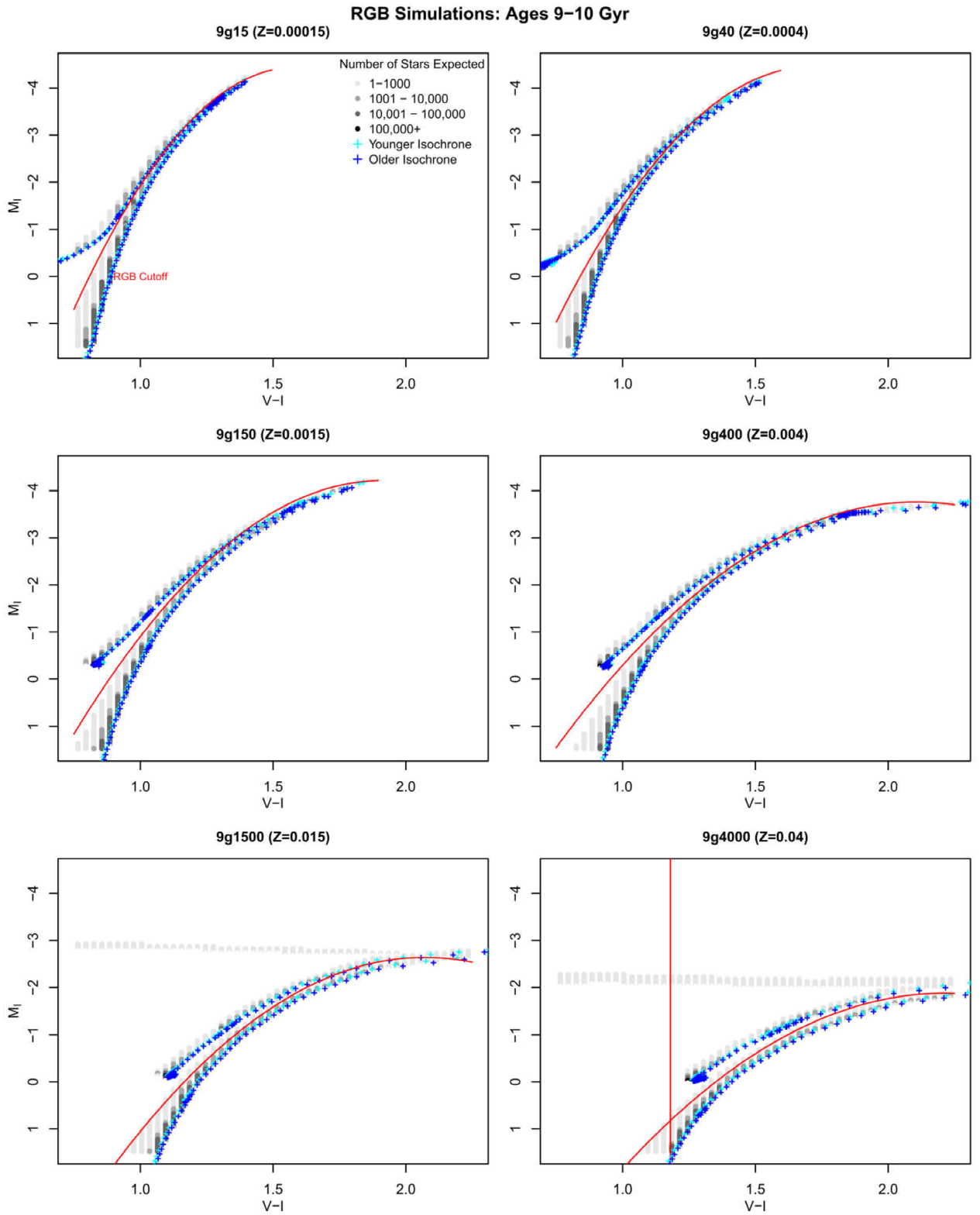}
	\caption{}
	\label{fig:app6}
\end{figure*}

\begin{figure*}
	\centering
	\includegraphics[scale=0.9]{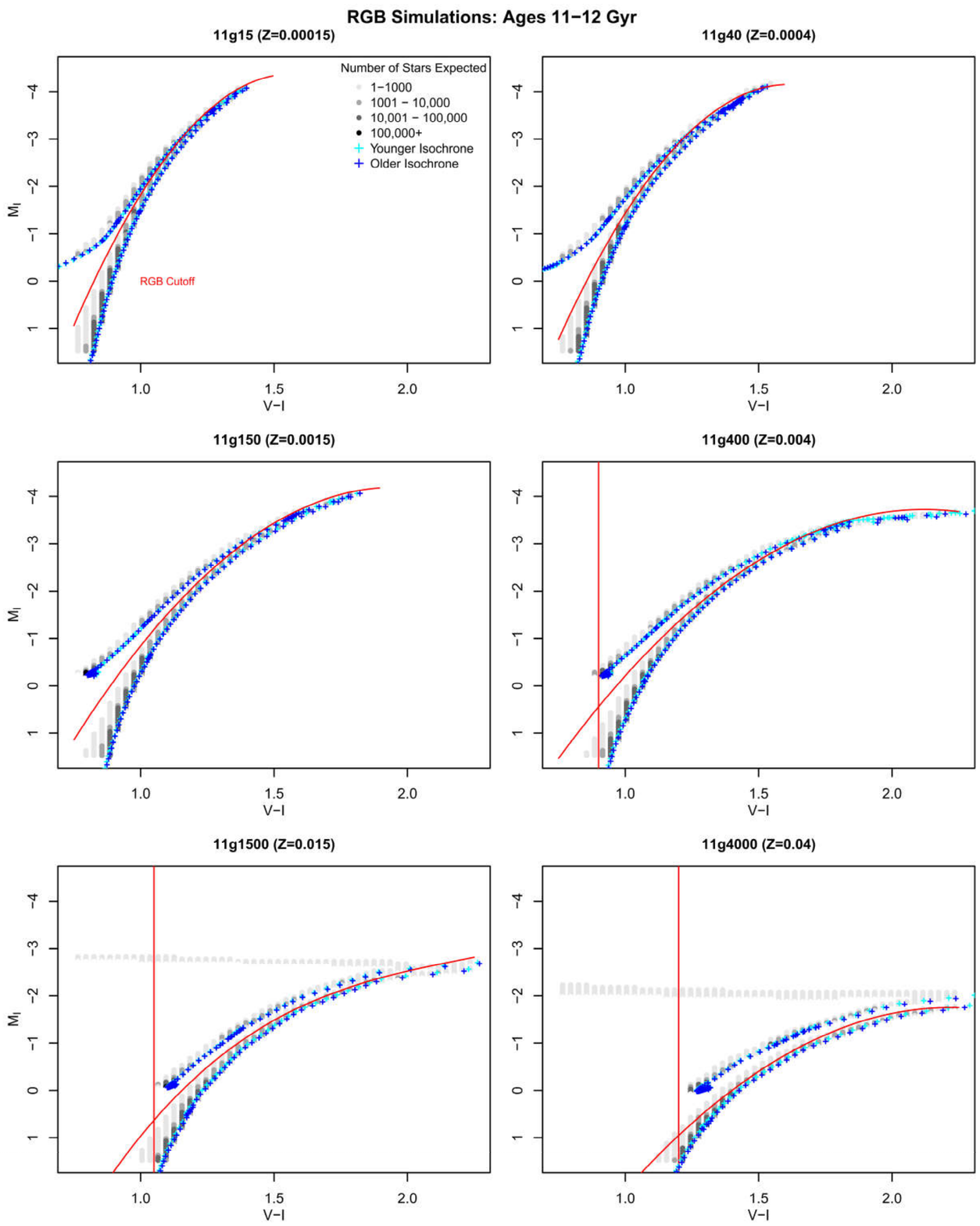}
	\caption{}
	\label{fig:app7}
\end{figure*}

\end{document}